# Improved three-dimensional thermal multiphase lattice Boltzmann model for liquid-vapor phase change

Qing Li[1,*], Y. Yu[1], and Kai H. Luo[2]

[1]*School of Energy Science and Engineering, Central South University, Changsha 410083, China*

[2]*Department of Mechanical Engineering, University College London, London WC1E 7JE, United Kingdom*

*Corresponding author: qingli@csu.edu.cn

## Abstract

Modeling liquid-vapor phase change using the lattice Boltzmann (LB) method has attracted significant attention in recent years. In this paper, we propose an improved three-dimensional (3D) thermal multiphase LB model for simulating liquid-vapor phase change. The proposed model has the following features. First, it is still within the framework of the thermal LB method using a temperature distribution function and therefore retains the fundamental advantages of the thermal LB method. Second, in the existing thermal LB models for liquid-vapor phase change, the finite-difference computations of the gradient terms $\nabla \cdot \mathbf{u}$ and $\nabla T$ usually require special treatment at boundary nodes, while in the proposed thermal LB model these two terms are calculated locally. Moreover, in some of the existing thermal LB models, the error term $\partial_{t_0}(T\mathbf{u})$ is eliminated by adding local correction terms to the collision process in the moment space, which causes these thermal LB models to be limited to the D2Q9 lattice in two dimensions and the D3Q15 or D3Q19 lattice in three dimensions. Conversely, the proposed model does not suffer from such an error term and therefore the thermal LB equation can be constructed on the D3Q7 lattice, which simplifies the model and improves the computational efficiency. Numerical simulations are carried out to validate the accuracy and efficiency of the proposed thermal multiphase LB model for simulating liquid-vapor phase change.

# I. Introduction

In the past three decades, the lattice Boltzmann (LB) method has been developed into an efficient numerical methodology for simulating fluid flow and heat transfer [1-6]. The fundamental idea of the LB method is to construct a simplified kinetic model that incorporates the essential physics of a target process so that the macroscopic averaged properties obey desired macroscopic equations [2]. Compared with traditional numerical methods based on the discretization of macroscopic continuum equations, the LB method exhibits some distinctive advantages owing to its kinetic nature, such as the simple form of the governing equation, the easiness of programming, and the easy implementation of complex boundary conditions. Besides, the LB method is much easier to parallelize due to its explicit scheme and local interactions. Furthermore, the LB method has been proven to be very suitable for modeling multiphase and multicomponent systems [4,6,7] where the interfacial dynamics and phase separation are present.

In recent years, the LB method has been extensively employed to simulate liquid-vapor phase change [4], such as boiling, condensation, and evaporation. Historically, the first LB study of boiling phenomena was attributed to Zhang and Chen in 2003 [8]. They successfully reproduced a nucleate boiling process by considering a standard Rayleigh-Bénard setup. Since then a variety of thermal multiphase LB models have been developed to simulate liquid-vapor phase change. Generally, these thermal multiphase LB models can be classified into three categories. The first category is based on the phase-field multiphase LB method, such as the models proposed by Dong *et al*. [9], Safari *et al*. [10,11], and Sun *et al*. [12]. In these models, the liquid-vapor interface is captured by solving an interface-capturing equation and a source term is incorporated into the continuity equation or the interface-capturing equation to mimic the liquid-vapor phase change [4]. The main weakness of these thermal multiphase LB models lies in that they are inefficient for simulating the nucleation of bubbles and seed bubbles should be prescribed *a priori* on a heated surface to trigger nucleate boiling.

The second category consists of some double-distribution-function (DDF) and hybrid thermal LB

models based on the pseudopotential multiphase LB method [13-16], such as the thermal multiphase LB models developed by Házi and Márkus [17,18], Biferale *et al*. [19,20], Gong and Cheng [21], Kamali *et al*. [22], Li *et al*. [23,24], and Zhang *et al*. [25]. An important advantage of the pseudopotential multiphase LB method is that the phase separation between different phases can emerge automatically as a result of particle interactions. Therefore, the interface between different phases can arise, deform, and migrate naturally, without resorting to any technique to track or capture the interface [13-15]. Accordingly, these thermal LB models are capable of simulating nucleate boiling without prescribing seed bubbles on a heated surface [23,26]. Most of the aforementioned thermal LB models utilize a temperature-based thermal LB equation to simulate the temperature field of non-ideal fluids. For the hybrid thermal LB model devised in Ref. [23], the temperature equation is solved by a finite-difference algorithm.

The third category is based on the multi-speed LB method, such as the thermal multiphase LB models proposed by Gonnella *et al*. [27] and Gan *et al*. [28]. These models employ a single density distribution function like that in the standard LB method but utilize more discrete velocities [29,30]. In these models, the equilibrium density distribution function includes higher-order velocity terms so as to recover the energy equation. Besides the aforementioned three categories, Reyhanian *et al*. [31] recently proposed a thermokinetic LB model for non-ideal fluids. In their model, the local thermodynamic pressure is imposed through appropriate rescaling of the particle's velocities and a total-energy-based thermal LB equation is employed to describe the energy conservation law. In addition, Huang *et al*. [32] recently devised a thermal multiphase LB model for liquid-vapor phase change by introducing a total-kinetic-energy-based thermal LB equation to recover the energy equation of non-ideal fluids.

Compared with the temperature-based thermal LB models mentioned in the second category, the total-energy-based and total-kinetic-energy-based thermal LB models usually have a simpler source term for simulating liquid-vapor phase change. Nevertheless, it should also be noted that the thermal boundary treatment becomes relatively complex in these models owing to the fact that the total energy or the total kinetic energy involves not only the temperature but also the density [31,32]. When a liquid-vapor interface meets a solid boundary with the Dirichlet thermal boundary condition, the temperature remains unchanged along the solid boundary but the density changes significantly. Correspondingly, the total energy and the total kinetic energy will also vary significantly along the solid boundary. In the LB community it is a little difficult to treat such a boundary condition with sufficient numerical stability for the related energy-based distribution functions over a wide range of applications and numerical instability may occur when complex boundaries are encountered.

Actually, the total-energy-based thermal LB method has been widely used in the simulations of single-phase incompressible thermal flows [33], which does not suffer from the aforementioned problem since the density is nearly constant ($\rho \approx \rho_0$) for single-phase incompressible thermal flows. Similarly, there is no such a problem in the thermal multiphase LB models that utilize a temperature distribution function, in which the thermal boundary treatment does not involve the density. This is also the main reason why many researchers prefer to use a temperature-based thermal multiphase LB model to simulate liquid-vapor phase change. However, the previous temperature-based thermal multiphase LB models [17-19,21,24,25] were criticized for the complex treatment of gradient terms such as $\nabla \cdot \mathbf{u}$ and $\nabla T$ at boundary nodes. Meanwhile, an error term caused by $\partial_{t_0}(T\mathbf{u})$ exists in some early models. To eliminate the error term, several improved 2D models [24,25] have been devised by adding correction terms to the collision process in the moment space. When these models are extended to the 3D space, the thermal LB equation would be limited to the D3Q15 or D3Q19 lattice [34].

In the present work, we aim at proposing an improved three-dimensional (3D) thermal multiphase LB model for simulating liquid-vapor phase change. To retain the advantages of the previous temperature-based thermal multiphase LB models, the present model is also a DDF-LB model and is constructed based on the pseudopotential multiphase LB method. To be specific, the density and velocity fields are simulated by a 3D pseudopotential multiphase LB model, while the temperature field is simulated by an improved thermal LB equation using a temperature distribution function. Meanwhile, the gradients $\nabla \cdot \mathbf{u}$ and $\nabla T$ in the source term can be calculated locally at boundary nodes in a simple way. Furthermore, the present model does not suffer from the error term caused by $\partial_{t_0}(T\mathbf{u})$ and therefore the thermal LB equation can be constructed on the D3Q7 lattice. The rest of this paper is organized as follows. A 3D pseudopotential multiphase LB model is briefly introduced in Sec. II. The improved 3D thermal multiphase LB model for liquid-vapor phase change is presented in Sec. III. Numerical validation is performed in Sec. IV. Finally, Sec. V summarizes the present paper.

## II. 3D pseudopotential multiphase LB model

The original pseudopotential multiphase LB method was proposed by Shan and Chen around 1993 [13,14]. This method has been applied with great success to many problems owing to its conceptual simplicity and computational efficiency [4]. In this section, we briefly introduce a 3D pseudopotential multiphase LB model. Using a multiple-relaxation-time (MRT) collision operator, the LB equation can be written as follows [35-38]:

$$f_\alpha\left(\mathbf{x} + \mathbf{e}_\alpha \delta_t,\, t + \delta_t\right) = f_\alpha\left(\mathbf{x}, t\right) - \overline{\Lambda}_{\alpha\beta}\left(f_\beta - f_\beta^{eq}\right)\Big|_{(\mathbf{x}, t)} + \delta_t\left(G_\alpha - 0.5\overline{\Lambda}_{\alpha\beta}G_\beta\right)\Big|_{(\mathbf{x}, t)},\qquad(1)$$

where $f_\alpha$ is the density distribution function, $f_\alpha^{eq}$ is the equilibrium density distribution function, $t$ is the time, $\mathbf{e}_\alpha$ is the discrete velocity along the $\alpha$-th direction, $\mathbf{x}$ is the spatial site, $\delta_t$ is the time step, $G_\alpha$ is the forcing term in the discrete velocity space, and $\overline{\Lambda}_{\alpha\beta} = \left(\mathbf{M}^{-1}\mathbf{\Lambda M}\right)$ is the collision operator, in which $\mathbf{M}$ is a transformation matrix and $\mathbf{\Lambda}$ is a diagonal matrix.

The D3Q19 lattice is adopted for the LB equation and the lattice velocities are given by

$$\mathbf{e}_\alpha = c\begin{bmatrix} 0 & 1 & -1 & 0 & 0 & 0 & 0 & 1 & -1 & 1 & -1 & 1 & -1 & 1 & -1 & 0 & 0 & 0 & 0 \\ 0 & 0 & 0 & 1 & -1 & 0 & 0 & 1 & -1 & -1 & 1 & 0 & 0 & 0 & 0 & 1 & -1 & 1 & -1 \\ 0 & 0 & 0 & 0 & 0 & 1 & -1 & 0 & 0 & 0 & 0 & 1 & -1 & -1 & 1 & 1 & -1 & -1 & 1 \end{bmatrix},\quad(2)$$

where $c = 1$ is the lattice constant. Multiplying Eq. (1) by the transformation matrix $\mathbf{M}$, the right-hand side of Eq. (1), i.e., the collision step of the LB equation, can be implemented in the moment space:

$$\mathbf{m}^* = \mathbf{m} - \mathbf{\Lambda}\left(\mathbf{m} - \mathbf{m}^{eq}\right) + \delta_t\left(\mathbf{I} - \frac{\mathbf{\Lambda}}{2}\right)\mathbf{S},\qquad(3)$$

where $\mathbf{m} = \mathbf{M}\mathbf{f}$ , $\mathbf{m}^{eq} = \mathbf{M}\mathbf{f}^{eq}$ , $\mathbf{I}$ is the unit matrix, and $\mathbf{S} = \mathbf{M}\mathbf{G}$ is the forcing term in the moment space. In the present work, the MRT collision operator formulated by Li *et al*. [37] is adopted. After the collision step in the moment space, $\mathbf{m}^{*}$ given by Eq. (3) can be transformed back to the discrete velocity space and then the streaming step is given by

$$f_{\alpha}\left(\mathbf{x} + \mathbf{e}_{\alpha}\delta_{t},\ t + \delta_{t}\right) = f_{\alpha}^{*}\left(\mathbf{x}, t\right), \qquad (4)$$

where $\mathbf{f}^{*} = \mathbf{M}^{-1}\mathbf{m}^{*}$, in which $\mathbf{M}^{-1}$ is the inverse matrix of the transformation matrix. The details of the transformation matrix $\mathbf{M}$ and its inverse matrix $\mathbf{M}^{-1}$ can be found in Ref. [37]. The corresponding equilibrium moments $\mathbf{m}^{eq} = \left(m_{0}^{eq}, m_{1}^{eq}, \cdots, m_{18}^{eq}\right)^{\mathrm{T}}$ in Eq. (3) are given by

$$m_{0}^{eq} = \rho, \quad m_{1}^{eq} = \rho u_{x}, \quad m_{2}^{eq} = \rho u_{y}, \quad m_{3}^{eq} = \rho u_{z}, \quad m_{4}^{eq} = \rho + \rho|\mathbf{u}|^{2},$$

$$m_{5}^{eq} = \rho\left(2u_{x}^{2} - u_{y}^{2} - u_{z}^{2}\right), \quad m_{6}^{eq} = \rho\left(u_{y}^{2} - u_{z}^{2}\right), \quad m_{7}^{eq} = \rho u_{x}u_{y}, \quad m_{8}^{eq} = \rho u_{x}u_{z}, \quad m_{9}^{eq} = \rho u_{y}u_{z},$$

$$m_{10}^{eq} = \rho c_{s}^{2}u_{y} + \rho u_{x}^{2}u_{y}, \quad m_{11}^{eq} = \rho c_{s}^{2}u_{x} + \rho u_{y}^{2}u_{x}, \quad m_{12}^{eq} = \rho c_{s}^{2}u_{z} + \rho u_{x}^{2}u_{z},$$

$$m_{13}^{eq} = \rho c_{s}^{2}u_{x} + \rho u_{z}^{2}u_{x}, \quad m_{14}^{eq} = \rho c_{s}^{2}u_{z} + \rho u_{y}^{2}u_{z}, \quad m_{15}^{eq} = \rho c_{s}^{2}u_{y} + \rho u_{z}^{2}u_{y},$$

$$m_{16}^{eq} = \varphi + \rho c_{s}^{2}\left(u_{x}^{2} + u_{y}^{2}\right), \quad m_{17}^{eq} = \varphi + \rho c_{s}^{2}\left(u_{x}^{2} + u_{z}^{2}\right), \quad m_{18}^{eq} = \varphi + \rho c_{s}^{2}\left(u_{y}^{2} + u_{z}^{2}\right), \qquad (5)$$

where $c_{s} = c/\sqrt{3}$ and $\varphi = \rho c_{s}^{4}\left(1 - 1.5|\mathbf{u}|^{2}\right)$, in which $c = 1$ and $|\mathbf{u}| = \sqrt{u_{x}^{2} + u_{y}^{2} + u_{z}^{2}}$. The first ten moments in Eq. (5) are related to the macroscopic density, momentum, and the viscous stress tensor, respectively, whereas the other moments are higher-order moments that do not affect the Navier-Stokes-level hydrodynamics. The diagonal matrix $\mathbf{\Lambda}$ for the relaxation times can be expressed as follows [37]:

$$\mathbf{\Lambda} = \mathrm{diag}\left(1, 1, 1, 1, \tau_{e}^{-1}, \tau_{v}^{-1}, \tau_{v}^{-1}, \tau_{v}^{-1}, \tau_{v}^{-1}, \tau_{v}^{-1}, \tau_{q}^{-1}, \tau_{q}^{-1}, \tau_{q}^{-1}, \tau_{q}^{-1}, \tau_{q}^{-1}, \tau_{q}^{-1}, \tau_{\pi}^{-1}, \tau_{\pi}^{-1}, \tau_{\pi}^{-1}\right), \qquad (6)$$

where $\tau_{e}$ and $\tau_{v}$ are determined by the bulk and shear viscosities, respectively, while $\tau_{q}$ and $\tau_{\pi}$ are free parameters related to high-order non-hydrodynamic moments. The relaxation times of the conserved moments have been chosen as 1.0 following Ref. [38].

According to the pseudopotential multiphase LB method [13,14], the intermolecular interaction force for single-component multiphase flows is given by [4,39]:

$$\mathbf{F}_m = -G\psi\left(\mathbf{x}\right)\sum_{\alpha}\omega\left(\left|\mathbf{e}_{\alpha}\right|^2\right)\psi\left(\mathbf{x}+\mathbf{e}_{\alpha}\delta_t\right)\mathbf{e}_{\alpha}, \tag{7}$$

where the weights $\omega\left(\left|\mathbf{e}_{\alpha}\right|^2\right)$ are given by $\omega(1)=1/6$ and $\omega(2)=1/12$ for the D3Q19 lattice, $G$ represents the strength of the interaction force, and the pseudopotential $\psi\left(\mathbf{x}\right)$ is defined as [40,41]

$$\psi\left(\mathbf{x}\right) = \sqrt{2\left[p_{\text{EOS}}\left(\rho,T\right)-\rho c_s^2\right]/Gc^2}, \tag{8}$$

where $p_{\text{EOS}}\left(\rho,T\right)$ is a non-ideal equation of state. In the present work, we adopt the Peng-Robinson equation of state [23,41]. Note that, when the pseudopotential is defined by Eq. (8), the only requirement for the parameter $G$ is to ensure that the whole term inside the square root on the right-hand side of Eq. (8) is positive and it is usually chosen as $G=-1$ in many practical applications. The forcing term $\mathbf{S}$ in Eq. (3) is given by [37]

$$\mathbf{S} = \begin{bmatrix} 0 \\ F_x \\ F_y \\ F_z \\ 2\mathbf{F}\cdot\mathbf{u}+\dfrac{6\sigma\left|\mathbf{F}_m\right|^2}{\psi^2\delta_t\left(\tau_e-0.5\right)} \\ 2\left(2F_x u_x - F_y u_y - F_z u_z\right) \\ 2\left(F_y u_y - F_z u_z\right) \\ F_x u_y + F_y u_x \\ F_x u_z + F_z u_x \\ F_y u_z + F_z u_y \\ c_s^2 F_y \\ c_s^2 F_x \\ c_s^2 F_z \\ c_s^2 F_x \\ c_s^2 F_z \\ c_s^2 F_y \\ 2c_s^2\left(u_x F_x + u_y F_y\right) \\ 2c_s^2\left(u_x F_x + u_z F_z\right) \\ 2c_s^2\left(u_y F_y + u_z F_z\right) \end{bmatrix}, \tag{9}$$

where $\sigma$ is a constant employed to adjust the mechanical stability condition of the pseudopotential LB model to achieve thermodynamic consistency [15,42]. The macroscopic density and velocity are calculated as follows:

$$\rho = \sum_\alpha f_\alpha, \quad \rho\mathbf{u} = \sum_\alpha \mathbf{e}_\alpha f_\alpha + \frac{\delta_t}{2}\mathbf{F}, \tag{10}$$

where $\mathbf{F}$ is the total force acting on the system. The Chapman-Enskog analysis of the aforementioned 3D pseudopotential multiphase LB model and the related numerical validation have been performed in Ref. [37].

## III. Improved 3D thermal LB model

In this section, an improved 3D thermal multiphase LB model is proposed based on the 3D pseudopotential LB model. Specifically, the density and velocity fields are simulated by the 3D pseudopotential LB model, while an improved thermal LB equation is devised to simulate the temperature field of non-ideal fluids. The coupling between the 3D pseudopotential LB model and the thermal LB equation is established via the non-ideal equation of state $p_{\text{EOS}}(\rho, T)$ in Eq. (8).

### A. Target macroscopic temperature equation

In 2002, He and Doolen [40] investigated the thermodynamic foundation of kinetic theory for multiphase flows and derived a macroscopic energy equation for non-ideal fluids with interfaces, which is given by (see Eq. (30) in Ref. [40]):

$$\partial_t(\rho e_n) + \nabla\cdot(\rho e_n\mathbf{u}) = \nabla\cdot(\lambda\nabla T) - \mathbf{P}:\nabla\mathbf{u} + \mathbf{\Pi}:\nabla\mathbf{u} + \kappa\left[\nabla(\rho\nabla\rho) - \frac{1}{2}\nabla\cdot(\rho\nabla\rho)\mathbf{I}\right]:\nabla\mathbf{u}, \tag{11}$$

where $\rho e_n$ is the internal energy density ($e_n$ is the internal energy), $\lambda$ is the thermal conductivity, $\mathbf{P}$ is the pressure tensor, $\mathbf{\Pi}$ is the viscous stress tensor, and $\kappa$ is the surface tension coefficient. The last term on the right-hand side of Eq. (11) stands for the heat generation by the surface tension. The internal energy $e_n$ includes the internal kinetic energy and the intermolecular potential energy [40], which can be expressed as follows:

$$\rho e_n = \rho e_n^0 - \frac{1}{2}\kappa\rho\nabla^2\rho, \tag{12}$$

where the superscript "0" is used to denote the standard bulk-phase thermodynamic properties of non-ideal fluids that do not involve interfaces [43]. Subsequently, Onuki [44,45] presented a dynamic van der Waals theory for liquid-vapor phase change. In their work the energy equation is expressed in terms of the total energy density $\rho e_n^T = \rho e_n + 0.5\rho |\mathbf{u}|^2$.

Using the thermodynamic relationships of non-ideal fluids, Eq. (11) can be transformed to the following temperature equation:

$$\rho c_V \left( \partial_t T + \mathbf{u} \cdot \nabla T \right) = \nabla \cdot \left( \lambda \nabla T \right) - T \left( \frac{\partial p_{EOS}}{\partial T} \right)_\rho \nabla \cdot \mathbf{u} + \mathbf{\Pi} : \nabla \mathbf{u} + \Phi , \qquad (13)$$

where $c_V$ is the specific heat at constant volume, $\mathbf{\Pi} : \nabla \mathbf{u}$ is the viscous heat dissipation, and the term $\Phi$ is proportional to the surface tension coefficient $\kappa$. In many previous studies [17-20,23-25], the last two terms on the right-hand side of Eq. (13) are neglected, which leads to the following equation:

$$\partial_t T + \mathbf{u} \cdot \nabla T = \frac{1}{\rho c_V} \nabla \cdot \left( \lambda \nabla T \right) - \frac{T}{\rho c_V} \left( \frac{\partial p_{EOS}}{\partial T} \right)_\rho \nabla \cdot \mathbf{u} . \qquad (14)$$

To match the thermal LB method, Eq. (14) is usually rewritten as follows [17-21,24,25]:

$$\partial_t T + \nabla \cdot \left( T \mathbf{u} \right) = \nabla \cdot \left( \chi \nabla T \right) + \phi , \qquad (15)$$

where $\phi$ is a source term. When $\chi$ in Eq. (15) is taken as the thermal diffusivity, i.e., $\chi = \lambda / (\rho c_V)$, the source term is given by [25]

$$\phi = \frac{\left( \chi \nabla T \right) \cdot \nabla \left( \rho c_V \right)}{\rho c_V} + T \left[ 1 - \frac{1}{\rho c_V} \left( \frac{\partial p_{EOS}}{\partial T} \right)_\rho \right] \nabla \cdot \mathbf{u} . \qquad (16)$$

It is noted that Eq. (15) is a convection-diffusion equation with a source term. However, in the literature it has been revealed that the standard thermal LB equation cannot recover a correct convection-diffusion equation [6,34] and it introduces an error term proportional to $\partial_{t_0} (T\mathbf{u})$ [4,6,24,34]. Such an error term can be negligible for incompressible thermal flows, but may lead to considerable numerical errors for multiphase flows [24]. In some of the previous studies [24,25], this error term is eliminated by adding local correction terms to the collision process in the moment space based on the MRT collision operator. Nevertheless, such treatment causes the thermal LB equation in these studies to be limited to the D2Q9 lattice. When this treatment is extended to the 3D space, the thermal LB equation would be limited to the D3Q15 or D3Q19 lattice [34].

Through the Chapman-Enskog analysis, it can be readily found that [24] the appearance of the error term proportional to $\partial_{t_0}(T\mathbf{u})$ is actually related to the recovery of the term $\nabla \cdot (T\mathbf{u})$ on the left-hand side of Eq. (15). In fact, from Eq. (16) we can see that the first term on the right-hand side of Eq. (16) can be written as $\mathbf{A} \cdot \nabla T$, which just takes the same form as the convective term $\mathbf{u} \cdot \nabla T$ on the left-hand side of Eq. (14). In other words, the target macroscopic temperature equation given by Eq. (14) can be rewritten as follows:

$$\partial_t T = \nabla \cdot (\chi \nabla T) + \phi_s , \tag{17}$$

where $\chi = \lambda/(\rho c_V)$ is the thermal diffusivity and the new source term $\phi_s$ is given by

$$\phi_s = \left( \frac{\chi \nabla (\rho c_V)}{\rho c_V} - \mathbf{u} \right) \cdot \nabla T - \frac{T}{\rho c_V} \left( \frac{\partial p_{\text{EOS}}}{\partial T} \right)_\rho \nabla \cdot \mathbf{u} . \tag{18}$$

Compared with the formulation given by Eqs. (15) and (16), the proposed new formulation given by Eqs. (17) and (18) has the following features. First, recovering the term $\nabla \cdot (T\mathbf{u})$ by the thermal LB equation is no longer needed. As a result, the error term caused by $\partial_{t_0}(T\mathbf{u})$ will disappear and the aforementioned limitation of the thermal LB models can be eliminated. Second, the transformation $\mathbf{u} \cdot \nabla T = \nabla \cdot (T\mathbf{u}) - T\nabla \cdot \mathbf{u}$ used in the formulation given by Eqs. (15) and (16) is also not needed in the new formulation. Recently, Zhang *et al*. [46] found that such transformation may lead to additional errors in numerical simulations.

In the following subsections, we will present an improved thermal LB equation that is aimed at recovering the temperature equation given by Eqs. (17) and (18). For practical applications, we prefer to use an MRT collision operator for the improved thermal LB equation since it has been well demonstrated that an MRT collision operator is superior over a Bhatnagar-Gross-Krook (BGK) collision operator in terms of numerical accuracy and stability [47]. Nevertheless, in order to provide a better understanding for general readers, we start with the BGK version of the improved thermal LB equation.

### B. Thermal LB-BGK equation

To recover the target macroscopic temperature equation formulated by Eqs. (17) and (18), the following thermal LB-BGK equation can be used:

$$g_\alpha \left( \mathbf{x} + \mathbf{e}_\alpha \delta_t, t + \delta_t \right) = g_\alpha \left( \mathbf{x}, t \right) - \frac{1}{\tau_g} \Big[ g_\alpha \left( \mathbf{x}, t \right) - g_\alpha^{eq} \left( \mathbf{x}, t \right) \Big] + \delta_t Q_\alpha \left( \mathbf{x}, t \right), \tag{19}$$

where $g_\alpha$ is the temperature distribution function, $g_\alpha^{eq} = w_\alpha T$ is the equilibrium temperature distribution function, $\tau_g$ is the non-dimensional relaxation time for the temperature field, and $Q_\alpha$ is the source term in the thermal LB equation. The macroscopic temperature is calculated by

$$T = \sum_\alpha g_\alpha. \tag{20}$$

For the D3Q7 lattice model, the discrete velocities are given by

$$\mathbf{e}_\alpha = \begin{bmatrix} 0 & 1 & -1 & 0 & 0 & 0 & 0 \\ 0 & 0 & 0 & 1 & -1 & 0 & 0 \\ 0 & 0 & 0 & 0 & 0 & 1 & -1 \end{bmatrix}. \tag{21}$$

Correspondingly, the weights of the equilibrium temperature distribution function $g_\alpha^{eq} = w_\alpha T$ can be chosen as $w_0 = 1 - d$ and $w_{1-6} = d/6$ ($d = 0.95$) for the D3Q7 lattice. Then $g_\alpha^{eq}$ satisfies

$$\sum_\alpha g_\alpha^{eq} = T, \quad \sum_\alpha \mathbf{e}_\alpha g_\alpha^{eq} = 0, \quad \sum_\alpha \mathbf{e}_\alpha \mathbf{e}_\alpha g_\alpha^{eq} = c_{sT}^2 T \mathbf{I}, \tag{22}$$

where $c_{sT} = \sqrt{d/3}$. For the D3Q15 and D3Q19 lattices, the weights $w_\alpha$ are the same as those used in the standard LB method and $c_{sT} = c_s = 1/\sqrt{3}$. In order to eliminate the discrete effect of the source term, $Q_\alpha \left( \mathbf{x}, t \right)$ in Eq. (19) is designed to be the following form:

$$Q_\alpha \left( \mathbf{x}, t \right) = \frac{3}{2} C_\alpha \left( \mathbf{x}, t \right) - \frac{1}{2} C_\alpha \left( \mathbf{x}, t - \delta_t \right), \tag{23}$$

in which $C_\alpha$ satisfies the following relationships:

$$\sum_\alpha C_\alpha = \phi_s, \quad \sum_\alpha \mathbf{e}_\alpha C_\alpha = 0, \tag{24}$$

where $\phi_s$ is given by Eq. (18). For simplicity, we can directly choose $C_\alpha = w_\alpha \phi_s$.

The Chapman-Enskog analysis can be employed to demonstrate that the macroscopic temperature equation formulated by Eqs. (17) and (18) can be correctly recovered from the thermal LB-BGK equation given by Eq. (19). Firstly, the following multi-scale expansions are introduced [48]:

$$\partial_t = \partial_{t_0} + \delta_t \partial_{t_1}, \quad g_\alpha = g_\alpha^{eq} + \delta_t g_\alpha^{(1)} + \delta_t^2 g_\alpha^{(2)}, \tag{25}$$

where $t_0$ and $t_1$ are two different time scales. The Taylor series expansion of Eq. (19) yields

$$\delta_t \left( \partial_t + \mathbf{e}_\alpha \cdot \nabla \right) g_\alpha + \frac{\delta_t^2}{2} \left( \partial_t + \mathbf{e}_\alpha \cdot \nabla \right)^2 g_\alpha + \cdots = -\frac{1}{\tau_g} \left( g_\alpha - g_\alpha^{eq} \right) + \delta_t Q_\alpha . \qquad (26)$$

Similarly, through the Taylor series expansion, $Q_\alpha$ given by Eq. (23) can be rewritten as

$$Q_\alpha \left( \mathbf{x}, t \right) = C_\alpha \left( \mathbf{x}, t \right) + \frac{1}{2} \delta_t \partial_t C_\alpha \left( \mathbf{x}, t \right) + O\left( \delta_t^2 \right) . \qquad (27)$$

Substituting Eq. (27) into Eq. (26) gives

$$\delta_t \left( \partial_t + \mathbf{e}_\alpha \cdot \nabla \right) g_\alpha + \frac{\delta_t^2}{2} \left( \partial_t + \mathbf{e}_\alpha \cdot \nabla \right)^2 g_\alpha + \cdots = -\frac{1}{\tau_g} \left( g_\alpha - g_\alpha^{eq} \right) + \delta_t C_\alpha + \frac{1}{2} \delta_t^2 \partial_t C_\alpha + \cdots . \qquad (28)$$

With the aid of Eq. (25), Eq. (28) can be written in the consecutive orders of $\delta_t$ as follows:

$$\delta_t : \quad \left( \partial_{t_0} + \mathbf{e}_\alpha \cdot \nabla \right) g_\alpha^{eq} = -\frac{1}{\tau_g} g_\alpha^{(1)} + C_\alpha , \qquad (29)$$

$$\delta_t^2 : \quad \partial_{t_1} g_\alpha^{eq} + \left( \partial_{t_0} + \mathbf{e}_\alpha \cdot \nabla \right) g_\alpha^{(1)} + \frac{1}{2} \left( \partial_{t_0} + \mathbf{e}_\alpha \cdot \nabla \right)^2 g_\alpha^{eq} = -\frac{1}{\tau_g} g_\alpha^{(2)} + \frac{1}{2} \partial_{t_0} C_\alpha . \qquad (30)$$

Substituting Eq. (29) into Eq. (30) leads to

$$\partial_{t_1} g_\alpha^{eq} + \left( \partial_{t_0} + \mathbf{e}_\alpha \cdot \nabla \right) \left( 1 - \frac{1}{2\tau_g} \right) g_\alpha^{(1)} + \frac{1}{2} \mathbf{e}_\alpha \cdot \nabla C_\alpha = -\frac{1}{\tau_g} g_\alpha^{(2)} . \qquad (31)$$

Taking the summations of Eqs. (29) and (31), the following equations can be obtained:

$$\partial_{t_0} T = \phi_s , \qquad (32)$$

$$\partial_{t_1} T + \nabla \cdot \left( 1 - \frac{1}{2\tau_g} \right) \left( \sum_\alpha \mathbf{e}_\alpha g_\alpha^{(1)} \right) = 0 . \qquad (33)$$

The relations $\sum_\alpha g_\alpha^{(1)} = \sum_\alpha g_\alpha^{(2)} = 0$, $\sum_\alpha C_\alpha = \phi_s$, and $\sum_\alpha \mathbf{e}_\alpha C_\alpha = 0$ together with $\sum_\alpha g_\alpha^{eq} = T$ and $\sum_\alpha \mathbf{e}_\alpha g_\alpha^{eq} = 0$ have been used in the above derivations. According to Eqs. (29) and (22), we have

$$\sum_\alpha \mathbf{e}_\alpha g_\alpha^{(1)} = -\tau_g \left[ \partial_{t_0} \left( \sum_\alpha \mathbf{e}_\alpha g_\alpha^{eq} \right) + \nabla \cdot \left( \sum_\alpha \mathbf{e}_\alpha \mathbf{e}_\alpha g_\alpha^{eq} \right) - \sum_\alpha \mathbf{e}_\alpha C_\alpha \right]$$
$$= -\tau_g c_{sT}^2 \nabla T . \qquad (34)$$

Combining Eq. (32) with Eqs. (33) and (34) through $\partial_t = \partial_{t_0} + \delta_t \partial_{t_1}$ gives

$$\partial_t T = \nabla \cdot \left( \chi \nabla T \right) + \phi_s , \qquad (35)$$

where $\chi = \lambda / \left( \rho c_V \right) = \left( \tau_g - 0.5 \right) c_{sT}^2 \delta_t$ and $\phi_s$ is given by Eq. (18).

Obviously, the above equation is exactly the target macroscopic temperature equation. It should be noted that the previous temperature-based thermal LB models yield $\sum_\alpha \mathbf{e}_\alpha g_\alpha^{eq} = T\mathbf{u}$, which leads to an error term proportional to $\partial_{t_0}(T\mathbf{u})$ according to Eq. (34). Such an error term may result in considerable numerical errors for liquid-vapor phase change [24]. In addition, from Eq. (34) it can be found that the temperature gradient in the source term $\phi_s$ can be calculated as follows:

$$\nabla T = -\frac{1}{\tau_g c_{sT}^2} \sum_\alpha \mathbf{e}_\alpha g_\alpha^{(1)} \approx -\frac{1}{\tau_g c_{sT}^2 \delta_t} \sum_\alpha \mathbf{e}_\alpha \left(g_\alpha - g_\alpha^{eq}\right). \tag{36}$$

A similar strategy can be found in the LB study of Li *et al.* [49] for axisymmetric thermal flows. In such a way, the temperature gradient in the source term can be calculated locally. Note that Eq. (36) utilized the approximation $\delta_t g_\alpha^{(1)} \approx g_\alpha - g_\alpha^{eq}$, which neglects the higher-order part $g_\alpha^{(2)}$ and therefore is invalid when the higher-order effects are dominant such as rarefied gas flows with a large Knudsen number.

### C. Thermal LB-MRT equation

Now we turn our attention to the MRT version of the improved thermal LB equation. Using an MRT collision operator, the thermal LB equation can be written as follows:

$$g_\alpha\left(\mathbf{x} + \mathbf{e}_\alpha \delta_t, t + \delta_t\right) = g_\alpha\left(\mathbf{x}, t\right) - \left(\mathbf{N}^{-1}\mathbf{\Gamma}\mathbf{N}\right)_{\alpha\beta}\left[g_\beta\left(\mathbf{x}, t\right) - g_\beta^{eq}\left(\mathbf{x}, t\right)\right] + \delta_t Q_\alpha\left(\mathbf{x}, t\right), \tag{37}$$

where $\mathbf{N}$ is a transformation matrix and $\mathbf{\Gamma}$ is a diagonal matrix for the relaxation times. For the D3Q15 and D3Q19 lattices, the transformation matrices used in the standard LB method can be directly adopted. For the D3Q7 lattice, the present work employs the following transformation matrix [50,51]:

$$\mathbf{N} = \begin{bmatrix} 1 & 1 & 1 & 1 & 1 & 1 & 1 \\ 0 & 1 & -1 & 0 & 0 & 0 & 0 \\ 0 & 0 & 0 & 1 & -1 & 0 & 0 \\ 0 & 0 & 0 & 0 & 0 & 1 & -1 \\ 0 & 1 & 1 & 1 & 1 & 1 & 1 \\ 0 & 1 & 1 & -1 & -1 & 0 & 0 \\ 0 & 1 & 1 & 0 & 0 & -1 & -1 \end{bmatrix}. \tag{38}$$

Through the transformation matrix $\mathbf{N}$, the first and second terms on the right-hand side of Eq. (37) can be executed in the moment space:

$$\mathbf{n}^* = \mathbf{n} - \mathbf{\Gamma}\left(\mathbf{n} - \mathbf{n}^{eq}\right), \tag{39}$$

where $\mathbf{n} = \mathbf{N}\mathbf{g}$ and $\mathbf{n}^{eq} = \mathbf{N}\mathbf{g}^{eq}$. Then the thermal LB-MRT equation can be rewritten as follows:

$$g_\alpha\left(\mathbf{x}+\mathbf{e}_\alpha\delta_t,\,t+\delta_t\right)=g_\alpha^*\left(\mathbf{x},\,t\right)+\delta_t Q_\alpha\left(\mathbf{x},t\right),\tag{40}$$

where $g^*=\mathbf{N}^{-1}\mathbf{n}^*$. Here $\mathbf{N}^{-1}$ is the inverse matrix of the transformation matrix $\mathbf{N}$ and is given by

$$\mathbf{N}^{-1}=\frac{1}{6}\begin{bmatrix}6 & 0 & 0 & 0 & -6 & 0 & 0\\0 & 3 & 0 & 0 & 1 & 1 & 1\\0 & -3 & 0 & 0 & 1 & 1 & 1\\0 & 0 & 3 & 0 & 1 & -2 & 1\\0 & 0 & -3 & 0 & 1 & -2 & 1\\0 & 0 & 0 & 3 & 1 & 1 & -2\\0 & 0 & 0 & -3 & 1 & 1 & -2\end{bmatrix}.\tag{41}$$

The equilibrium temperature distribution function is still given by $g_\alpha^{eq}=w_\alpha T$, in which the weights $w_\alpha$ are chosen as $w_0=1-d$ and $w_{1-6}=d/6$ ($d=0.95$) for the D3Q7 lattice. Using the relation $\mathbf{n}^{eq}=\mathbf{N}g^{eq}$, the equilibrium moments $\mathbf{n}^{eq}=\left(n_0^{eq},n_1^{eq},\cdots,n_6^{eq}\right)^{\mathrm{T}}$ can be obtained as follows:

$$\mathbf{n}^{eq}=\left(T,\,0,\,0,\,0,\,3c_{sT}^2 T,\,0,\,0\right)^{\mathrm{T}},\tag{42}$$

where $c_{sT}=\sqrt{d/3}$. Correspondingly, the diagonal matrix $\mathbf{\Gamma}$ for the relaxation times is given by

$$\mathbf{\Gamma}=\mathrm{diag}\left(1,\,\varsigma_T^{-1},\,\varsigma_T^{-1},\,\varsigma_T^{-1},\,\varsigma_q^{-1},\,\varsigma_\pi^{-1},\,\varsigma_\pi^{-1}\right),\tag{43}$$

where $\varsigma_T$ is the relaxation time related to the thermal diffusivity, while $\varsigma_q$ and $\varsigma_\pi$ are free parameters.

By conducting a similar Chapman-Enskog analysis for the aforementioned thermal LB-MRT equation, it can be readily verified that the macroscopic temperature equation formulated by Eq. (17) can be correctly recovered and the thermal diffusivity is given by $\chi=\lambda/\left(\rho c_V\right)=\left(\varsigma_T-0.5\right)c_{sT}^2\delta_t$. Meanwhile, the temperature gradient can be calculated locally as follows:

$$\nabla T\approx-\frac{1}{\varsigma_T c_{sT}^2\delta_t}\sum_\alpha\mathbf{e}_\alpha\left(g_\alpha-g_\alpha^{eq}\right).\tag{44}$$

The above equation is just the same as Eq. (36) except that the single relaxation time $\tau_g$ has been replaced by the relaxation time $\varsigma_T$. Besides, Eq. (44) can also be written as follows:

$$\partial_x T\approx-\frac{1}{\varsigma_T c_{sT}^2\delta_t}\left(n_1-n_1^{eq}\right),\quad\partial_y T\approx-\frac{1}{\varsigma_T c_{sT}^2\delta_t}\left(n_2-n_2^{eq}\right),\quad\partial_z T\approx-\frac{1}{\varsigma_T c_{sT}^2\delta_t}\left(n_3-n_3^{eq}\right).\tag{45}$$

This equation can be applied in the whole computational domain for computing the temperature gradient in the source term.

Moreover, according to the Chapman-Enskog analysis of the aforementioned 3D pseudopotential LB model [37], the following equation can be obtained:

$$\partial_{t_0} m_4^{eq} + \partial_x \left( m_1^{eq} + m_{11}^{eq} + m_{13}^{eq} \right) + \partial_y \left( m_2^{eq} + m_{10}^{eq} + m_{15}^{eq} \right) + \partial_z \left( m_3^{eq} + m_{12}^{eq} + m_{14}^{eq} \right)$$
$$= -\frac{1}{\tau_e} m_4^{(1)} + \left( 1 - \frac{1}{2\tau_e} \right) S_4, \tag{46}$$

where $\delta_t m_4^{(1)} \approx m_4 - m_4^{eq}$, $\tau_e$ can be found in Eq. (6), and $S_4$ is the fifth moment of the forcing term **S** in Eq. (9). Substituting the equilibrium moments given by Eq. (5) into Eq. (46) leads to

$$2\rho c_s^2 \boldsymbol{\nabla} \cdot \mathbf{u} + 2\mathbf{F} \cdot \mathbf{u} = -\frac{1}{\tau_e} m_4^{(1)} + \left( 1 - \frac{1}{2\tau_e} \right) S_4 + 3\rho u_i^2 \partial_i u_i + u_i^3 \partial_i \rho, \tag{47}$$

where $i$ is a dummy summation index. The last two terms on the right-hand side of Eq. (47) are cubic velocity terms. In previous studies, the velocity gradient is usually calculated via an isotropic finite-difference scheme (see Eq. (73) in Ref. [4]), which can be well implemented for internal fluid nodes (see Fig. 1), but is not suitable for boundary nodes near solid walls. Therefore special treatment is usually required for the boundary nodes. Actually, since the velocity near a non-slip solid wall is very small, the cubic velocity terms in Eq. (47) can be neglected and the velocity divergence at boundary nodes can be approximately calculated by

$$\boldsymbol{\nabla} \cdot \mathbf{u} \approx \frac{1}{2\rho c_s^2} \left[ -\frac{1}{\tau_e \delta_t} \left( m_4 - m_4^{eq} \right) + \left( 1 - \frac{1}{2\tau_e} \right) S_4 - 2\mathbf{F} \cdot \mathbf{u} \right]. \tag{48}$$

In summary, Eqs. (39), (40), and (23) together with Eqs. (45) and (48) constitute the improved thermal LB-MRT equation for simulating the temperature field of non-ideal fluids.

Finally, we would like to mention the wetting boundary scheme used in the present work. A two-dimensional sketch of the grid nodes near a solid wall is shown in Fig. 1. The non-slip solid wall is located at the middle of the fluid boundary nodes and the solid nodes. In this work, we utilize an improved virtual-density contact-angle scheme [52] to implement the wetting boundary, which specifies a virtual density for a solid node. As a result, the density gradient $\boldsymbol{\nabla}\rho$ at a fluid boundary node can be well calculated by the isotropic finite-difference scheme used in the multiphase LB community [4].

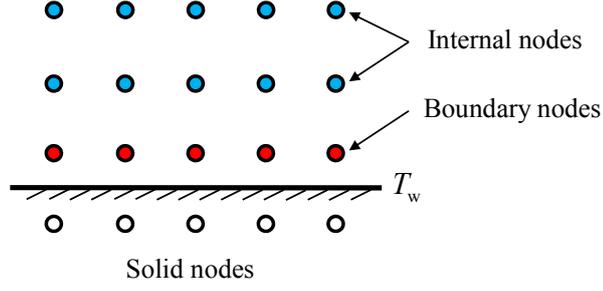

**Fig. 1**. Two-dimensional sketch of the grid nodes near a solid wall. The blue nodes are internal fluid nodes, the red nodes are boundary nodes, while the nodes beneath the solid wall are solid nodes.

## IV. Numerical simulations

In the previous section, an improved 3D thermal multiphase LB model has been proposed for liquid-vapor phase change, which consists of a 3D pseudopotential LB model for simulating the density and velocity fields, and an improved thermal LB equation for modeling the temperature field. In this section, numerical simulations are carried out to validate the proposed model. Unless otherwise mentioned, the MRT version of the improved thermal LB equation is used and the relaxation times in Eq. (43) are all taken as 1.0 except that $\varsigma_T$ is determined by the thermal diffusivity. Three tests are considered. Firstly, the well-known $D^2$ law for droplet evaporation is employed to validate the proposed model. Considering that the test of $D^2$ law does not involve solid boundaries, we later simulate droplet evaporation on a heated surface, and finally the test of bubble nucleation and departure is utilized to verify the capability of the proposed model for simulating nucleate boiling.

### A. Validation of the $D^2$ law

The $D^2$ law states that the square of the diameter of an evaporating droplet decreases linearly with time, i.e., $D^2(t)/D_0^2 = 1 - kt$ [53,54]. This law is mainly established based on the following assumptions: the evaporation occurs in an environment with negligible viscous heat dissipation and no buoyancy, the liquid and vapor phases are quasi-steady, and the thermo-physical properties are constant. Numerical simulations are carried out in a 3D cubic domain $L_x \times L_y \times L_z = 100\,\text{l.u.} \times 100\,\text{l.u.} \times 100\,\text{l.u.}$, where l.u. denotes the lattice units. Initially, a droplet with a diameter of $D_0 = 50\,\text{l.u.}$ is located at the center of the computational domain.

In this work, we adopt the Peng-Robinson equation of state [41]:

$$p_{\text{EOS}} = \frac{\rho R T}{1 - b\rho} - \frac{a \vartheta(T) \rho^2}{1 + 2b\rho - b^2 \rho^2},\qquad(49)$$

where $\vartheta(T) = \left[ 1 + \left( 0.37464 + 1.54226\omega - 0.26992\omega^2 \right)\left( 1 - \sqrt{T/T_c} \right) \right]^2$ with $\omega = 0.344$ being the acentric factor, $a = 0.45724 R^2 T_c^2 / p_c$, and $b = 0.0778 R T_c / p_c$, in which $R$ is the gas constant, $T_c$ is the critical temperature, and $p_c$ is the critical pressure. In simulations, $\left( \partial p_{\text{EOS}} / \partial T \right)_\rho$ is taken as $\left( \partial p_{\text{EOS}} / \partial T \right)_\rho \approx \rho R / (1 - b\rho)$ and the parameters are chosen as $a = 1/49$, $b = 2/21$, and $R = 1$. The saturation temperature of the system is chosen as $T_{\text{sat}} = 0.86 T_c$, which corresponds to $\rho_l \approx 6.5$ and $\rho_g \approx 0.38$. Initially, the temperature of the droplet is $T_{\text{sat}}$ and the temperature of the surrounding vapor is given by $T_g = T_{\text{sat}} + \Delta T$, in which the superheat $\Delta T$ is taken as $0.14 T_c$. The specific heat at constant volume is chosen as $c_V = 5.0$ [24]. The kinematic viscosity is fixed at $\nu = 0.1$ for both the liquid and vapor phases. Accordingly, the relaxation time $\tau_\nu$ in Eq. (6) is given by $\tau_\nu = 0.8$. The relaxation time $\tau_e$ is chosen as $\tau_e = 1.25$, while the remaining relaxation times are taken as 1.0. The spurious currents are examined via the simulation of a static droplet at the saturation temperature and the maximum spurious current is found to be about $7.0 \times 10^{-4}$.

At the boundaries, a constant temperature ($T_g$) is applied. According to the assumption of the $D^2$ law, the thermal conductivity $\lambda$ is constant. In this test, $\lambda$ is taken as 0.2. Figure 2 shows some snapshots of the density contours obtained by the present thermal LB model and those given by a hybrid thermal LB model, which employs a finite-difference algorithm (FDM) [23] to solve the macroscopic temperature equation given by Eq. (14). Owing to the temperature gradient at the liquid-vapor interface, the droplet gradually evaporates. Quantitatively, the variations of $\left( D/D_0 \right)^2$ with time are displayed in Fig. 3. For comparison, the results obtained by a 3D thermal LB equation devised by Biferale *et al.* [19,20] are also presented in the figure. The linear relationship between $\left( D/D_0 \right)^2$ and the time can be observed in Fig. 3 for the results given by the present model and those obtained by a finite-difference algorithm for solving the macroscopic temperature equation and there are no significant differences between them. However, the thermal LB model of Biferale *et al*. [19,20], which suffers from the error term $\partial_{t_0}(T\mathbf{u})$, leads to considerable numerical errors as shown in Fig. 3.

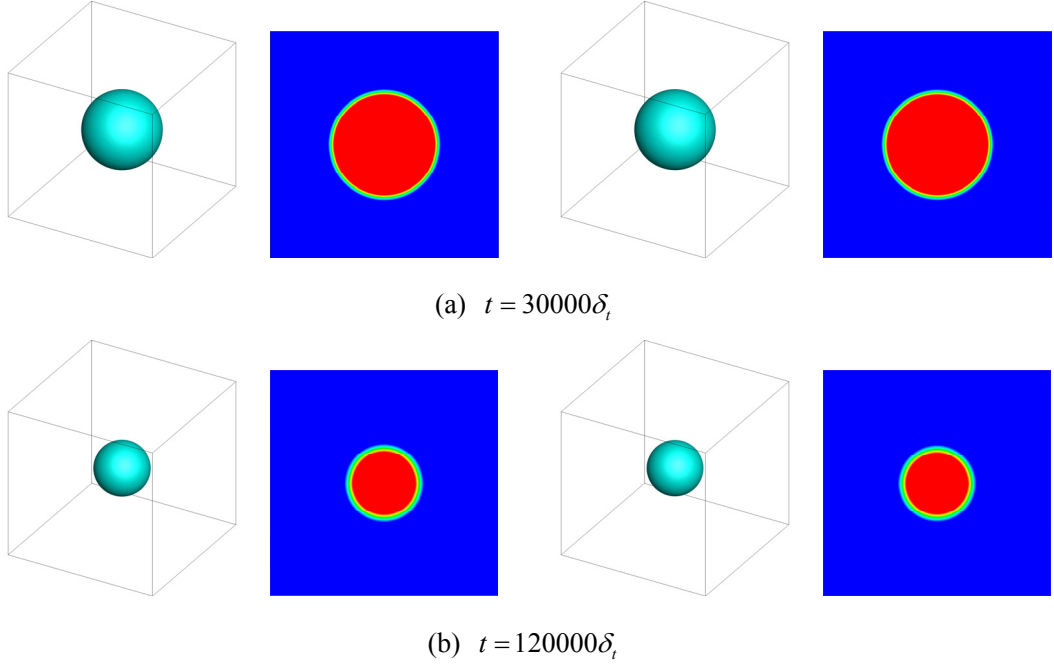

(a) $t = 30000\delta_t$

(b) $t = 120000\delta_t$

**Fig. 2**. Numerical validation of D$^2$ law. Snapshots of the density contours obtained by the present thermal LB model (left) and a hybrid thermal LB model with a finite-difference algorithm (right).

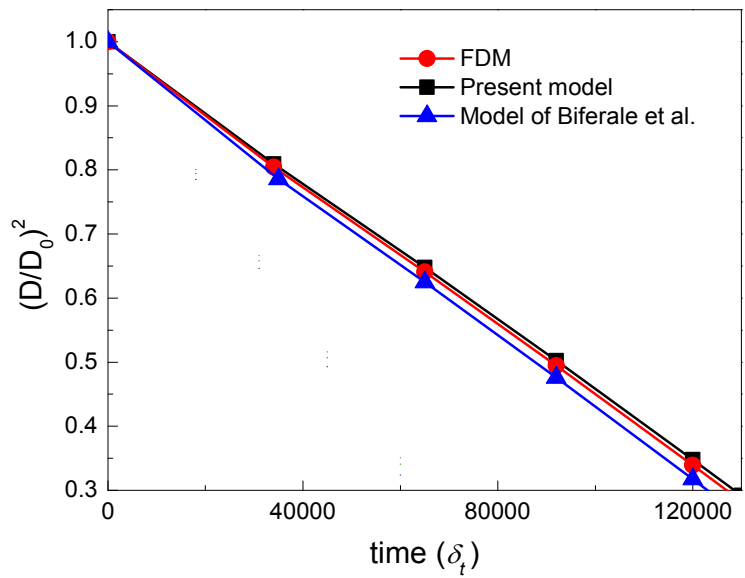

**Fig. 3**. Numerical validation of D$^2$ law. The variations of $\left(D/D_0\right)^2$ with time.

## B. Droplet evaporation on a heated surface

The preceding test does not involve solid boundaries. To further validate the capability of the proposed thermal LB model, in this subsection we consider the test of droplet evaporation on a heated solid surface. The grid system is chosen as $L_x \times L_y \times L_z = 200 \, \text{l.u.} \times 200 \, \text{l.u.} \times 100 \, \text{l.u.}$ with the periodic boundary condition being applied in the $x$ and $y$ directions. A heated surface is located at the bottom of the computational domain. Initially, a semi-spherical droplet with a radius of $r = 30 \, \text{l.u.}$ is placed at the center of the bottom surface. The equilibrium contact angle of the bottom surface is chosen as $\theta \approx 88^{\circ}$ [52]. The popular halfway bounce-back scheme [55] is utilized to treat the unknown density distribution functions at the boundary nodes around the bottom solid surface (see Fig. 1):

$$f_{\bar{\alpha}}\left(\mathbf{x}_{b}, t + \delta_{t}\right) = f_{\alpha}^{*}\left(\mathbf{x}_{b}, t\right), \tag{50}$$

where $\mathbf{x}_{b}$ denotes a fluid boundary node and $\mathbf{e}_{\bar{\alpha}} = -\mathbf{e}_{\alpha}$. Similarly, the anti-bounce-back scheme [56] is adopted to treat the Dirichlet thermal boundary condition:

$$g_{\bar{\alpha}}\left(\mathbf{x}_{b}, t + \delta_{t}\right) = -g_{\alpha}^{+}\left(\mathbf{x}_{b}, t\right) + 2w_{\alpha}T_{w}, \tag{51}$$

where $g_{\alpha}^{+}\left(\mathbf{x}_{b}, t\right)$ represents the right-hand side of Eq. (40).

The main simulation parameters are the same as those used in the previous test except that the thermal conductivity is taken as $\lambda = \rho c_{V} \chi$ with $\chi = 0.02$ [24]. The saturation temperature is still taken as $T_{sat} = 0.86 T_{c}$, while the temperature of the heated surface is chosen as $T_{w} = 0.87 T_{c}$. The open boundary condition is employed at the top of the domain with the temperature being fixed at $T_{sat}$. The first $10^{4}$ time steps of the simulations are carried out without evaporation so that the droplet can reach its equilibrium state on the bottom surface and the thermal LB model is added after $t = 10^{4} \delta_{t}$. During the evaporation process, the contact angle hysteresis [57] is taken into consideration with a hysteresis window of $\left(0^{\circ}, 180^{\circ}\right)$. As a result, the droplet will evaporate on the heated bottom surface in the Constant-Contact-Radius (CCR) mode. Figure 4 illustrates the variations of the normalized droplet mass with time during the evaporation process. As shown in the figure, the numerical results predicted by the present thermal LB model are in good agreement with those obtained by solving the macroscopic temperature equation with a finite-difference algorithm, while the thermal LB model of Biferale *et al.* [19,20] yields considerable numerical errors. Some snapshots of the density contours obtained by the present model are displayed in Fig. 5, from which the CCR evaporation mode can be clearly observed.

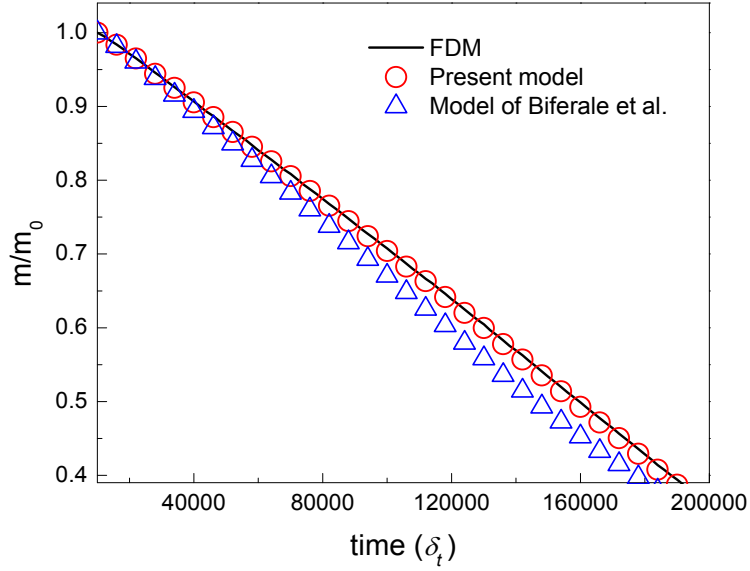

**Fig. 4**. Variations of normalized droplet mass with time during droplet evaporation on a heated surface.

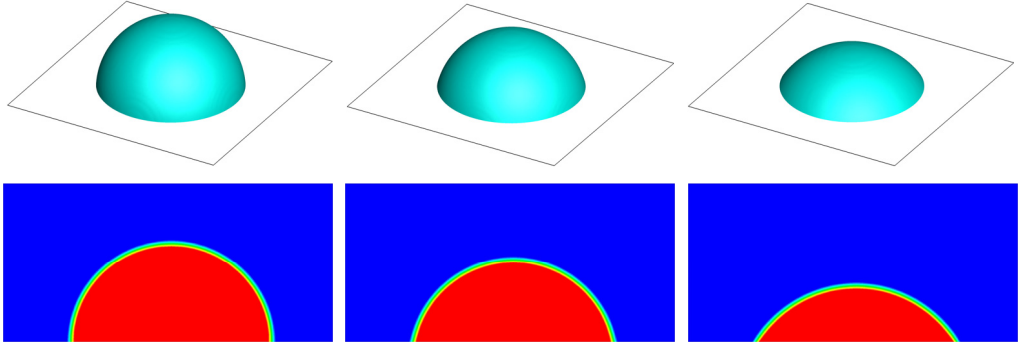

**Fig. 5**. Snapshots of the density contours obtained by the present model during droplet evaporation on a heated surface. From left to right: $t = 10000\delta_t$, $90000\delta_t$, and $180000\delta_t$, respectively. A 3D view is shown in the top row with $x \in [50, 150]$, $y \in [50, 150]$, and $z \in [0, 50]$, while in the bottom row the density contours of the $y$-$z$ cross section at $x = 100$ are presented.

## C. Bubble nucleation and departure

In this subsection, numerical simulations are performed for bubble nucleation and departure to verify the capability of the proposed thermal LB model for simulating nucleate boiling. The grid system of the computational domain is taken as $L_x \times L_y \times L_z = 200 \, \text{l.u.} \times 200 \, \text{l.u.} \times 280 \, \text{l.u.}$. The kinematic viscosity, the specific heat at constant volume, the saturation temperature, and the relaxation parameters are the same as those used in the previous subsection. The thermal conductivity is given by $\lambda = \rho c_V \chi$ with $\chi = 0.08$. The parameters of the equation of state are taken as $a = 2/49$, $b = 2/21$, and $R = 1$. Initially, the domain is filled with saturated liquid $(0 \leq z \leq 200 \, \text{l.u.})$ below its vapor at $T_{\text{sat}} = 0.86T_c$. The temperature of the bottom solid surface is fixed at $T_w = 0.96T_c$. A square hydrophobic region with its side length of 55 l.u. is located at the center of the bottom surface and the equilibrium contact angle is set to $\theta_{\text{pho}} \approx 130°$. The rest region of the bottom surface is hydrophilic with $\theta_{\text{phi}} \approx 37°$. The halfway bounce-back scheme is employed at the boundary nodes around the top and bottom surfaces, while the periodic boundary condition is applied in the $x$ and $y$ directions. A buoyant force is applied in the computational domain, i.e., $\mathbf{F}_b = (\rho - \rho_{\text{ave}})\mathbf{g}$, where $\rho_{\text{ave}}$ is the average density over the domain and $\mathbf{g} = (0, \, 0, \, -g)$ is the gravitational acceleration.

Figure 6 displays some snapshots of the density contours obtained by the present thermal LB model and a hybrid thermal LB model with a finite-difference algorithm in the case of $g = 1.5 \times 10^{-5}$. From Fig. 6(a) we can see that a bubble is nucleated at the central region of the bottom surface, which arises from the hydrophobicity of the central region [26]. As time goes by, the bubble gradually grows up. After reaching its departure diameter, the bubble is separated into two parts, i.e., a departure bubble and a residual bubble left on the bottom surface, as shown in Fig. 6(c). Generally, it can be seen that the bubble nucleation, growth, and departure processes are well simulated and there are no obvious differences between the bubble dynamic behaviors predicted by the present thermal LB model and those produced by a hybrid thermal LB model that utilizes a finite-difference algorithm to solve the macroscopic temperature equation.

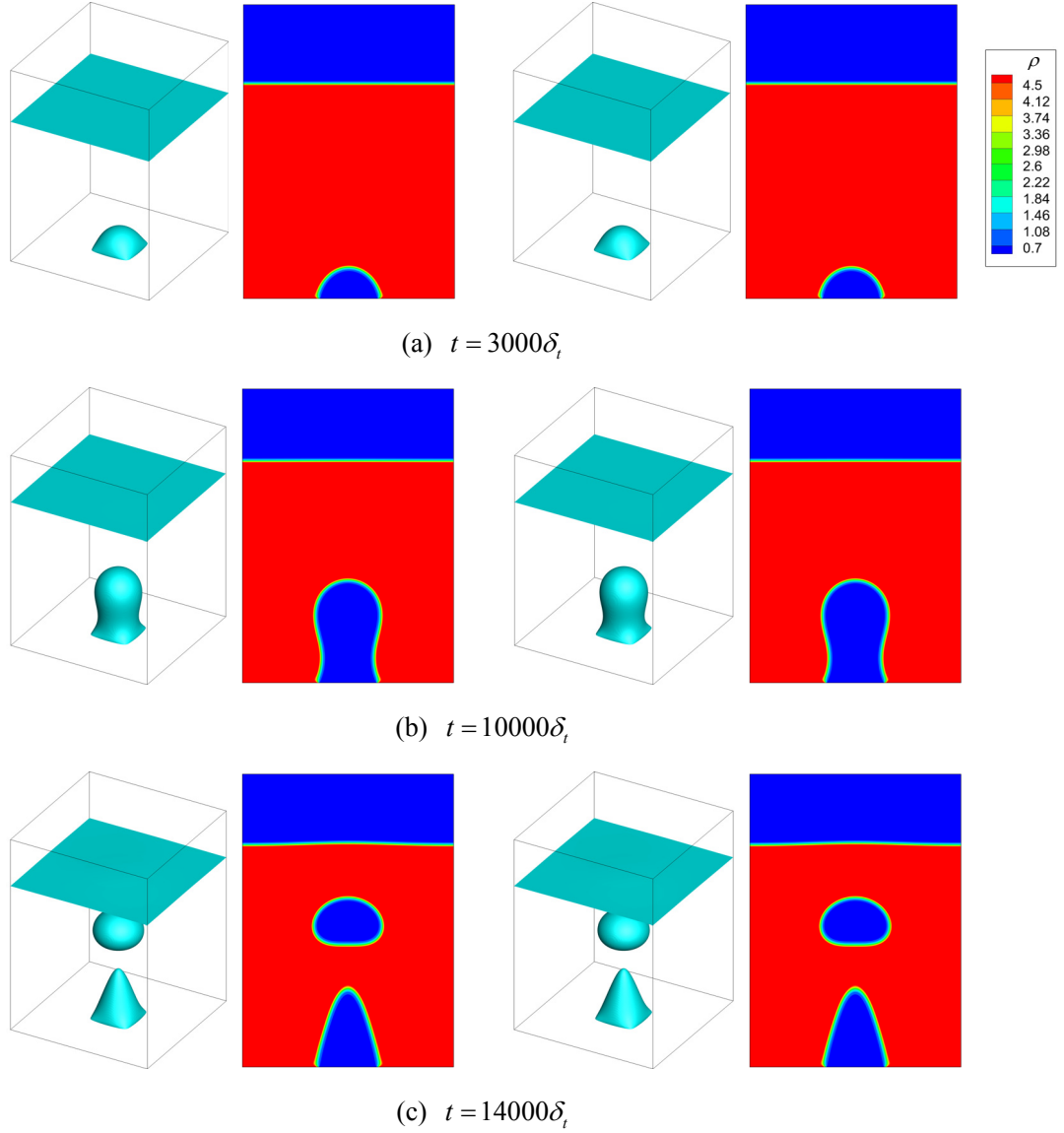

(a)  $t = 3000\delta_t$

(b)  $t = 10000\delta_t$

(c)  $t = 14000\delta_t$

**Fig. 6**. Simulations of bubble nucleation and departure. Snapshots of the density contours obtained by the present thermal LB model (left) and a hybrid thermal LB model with a finite-difference algorithm (right) in the case of $g = 1.5 \times 10^{-5}$. The snapshots include a 3D view and the cross-sectional view at $x = 100$.

Furthermore, to quantify the numerical results, the power-law relationship between the bubble departure diameter and the gravitational acceleration is verified. Figure 7 illustrates the variation of the bubble departure diameter with the gravitational acceleration, in which the solid line represents the results of $D_d = 0.253 g^{-0.5}$. From the figure we can see that the values of the bubble departure diameter predicted by the present thermal LB model are basically in good agreement with those given by the solid line, which indicates that the results obtained by the present model are consistent with the experimental correlations in the literature [58,59] (i.e., $D_d \sim g^{-0.5}$). Furthermore, no significant differences are observed in Fig. 7 between the results of the present thermal LB model and those given by a hybrid thermal LB model with a finite-difference algorithm.

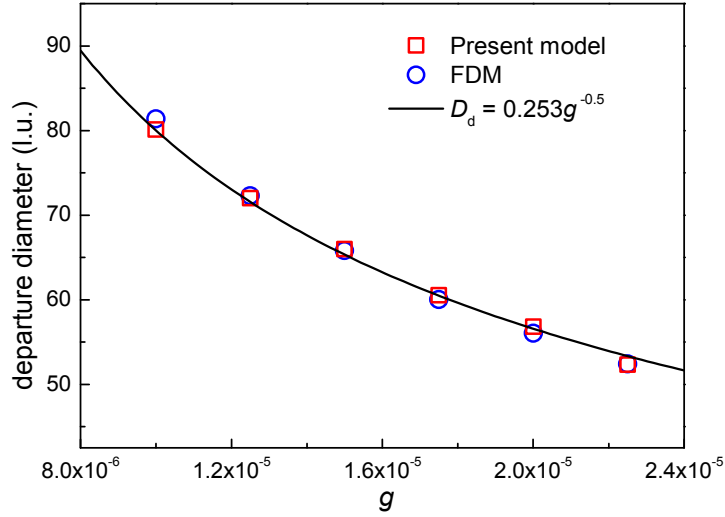

**Fig. 7**. Simulations of bubble nucleation and departure. Numerical validation of the power-law relationship between the bubble departure diameter and the gravitational acceleration.

# V. Summary

In this paper, we have developed an improved 3D thermal multiphase LB model for simulating liquid-vapor phase change, which consists of a 3D pseudopotential LB model for simulating the density and velocity fields and an improved thermal LB equation for modeling the temperature field of non-ideal fluids. The coupling between the 3D pseudopotential LB model and the thermal LB equation is established via the non-ideal equation of state. The improved model does not suffer from the error term caused by $\partial_{t_0}(T\mathbf{u})$ and therefore the thermal LB equation can be constructed on the D3Q7 lattice to simplify the model and improve the computational efficiency. Moreover, in the previous temperature-based thermal LB models for liquid-vapor phase change, the finite-difference computations of the gradient terms $\nabla \cdot \mathbf{u}$ and $\nabla T$ usually require special treatment at boundary nodes, while in the improved model these two terms can be calculated locally in a simple way. The accuracy and efficiency of the improved thermal multiphase LB model for simulating liquid-vapor phase change have been well validated by testing the well-known $D^2$ law, droplet evaporation on a heated surface, and bubble nucleation and departure in nucleate boiling.

In the literature, most of the previous LB studies on liquid-vapor phase change were limited to the so-called wet-node system, in which the solid boundaries are located on lattice nodes. In contrast, the aforementioned advantages of the proposed thermal multiphase LB model make it suited not only for the wet-node system but also for the link-wise system (see Fig. 1), which is often utilized with the popular halfway bounce-back scheme to simulate fluid flows in complex geometries, such as porous media. Furthermore, the proposed model may be useful for thermal non-equilibrium flows [60] and can be extended to the problems involving curved boundaries by combining it with the related hydrodynamic and thermal curved boundary schemes in the LB community [61,62]. Finally, it is worth mentioning that the proposed thermal LB equation can also be used on the D3Q15 or D3Q19 lattice for both the BGK and MRT collision operators. A major advantage of the MRT version lies in its better numerical stability than the BGK version as the thermal diffusivity varies over a wide range.

## Acknowledgments

This work was supported by the National Natural Science Foundation of China (No. 52176093 and No. 51822606).

Scientific, 2013).